\documentstyle[11pt,twoside,epsfig,rotating,amstex,cite]{article}

\newcommand{\beq}{\begin{equation}}                                            
\newcommand{\eeq}{\end{equation}}

\newlength{\dinwidth}                                                          
\newlength{\dinmargin}                                                         
\begin{document}                                                               
\noindent                                                                      
\begin{titlepage}                                                              
\begin{flushleft}
%
%
{\tt DESY 97-042    \hfill    ISSN 0418-9833} \\
{\tt March 1997}                  \\
\end{flushleft}
                                                                               
\vspace*{3.cm}                                                                 
\begin{center}                                                                 
\begin{LARGE}                                                                  
{\bf  A Measurement  of the
Proton Structure Function {\boldmath $F_2(x,Q^2)$} 
at Low {\boldmath $x$} and Low {\boldmath $Q^2$}
at HERA\\ }                           
\vspace*{2.cm}                                                                 
  H1 Collaboration \\                                                          
\end{LARGE}                                                                    
\vspace*{4.cm}                                                                 
{\bf Abstract:}                                                                
\begin{quotation}                                                              
\noindent
The results of                                                       
a measurement of the proton structure function $F_2(x,Q^2)$ 
and the virtual photon-proton cross section are 
reported       
for momentum transfers squared $Q^2$ between 0.35~GeV$^2$ and
3.5~GeV$^2$ and for Bjorken-$x$ values down to $6\cdot 10^{-6}$           
using data collected by the HERA experiment H1 in 1995.                        
The data represent an increase in kinematic reach to lower
$x$ and $Q^2$ values of about a factor of 5 compared to 
previous H1 measurements. Including   measurements from
fixed target experiments
the rise of $F_2$ with decreasing $x$ is found to be less steep 
for the lowest $Q^2$ values measured.
Phenomenological models at low $Q^2$                               
are compared with the data.
                          
\end{quotation}                                                                
\vfill                                                                         
\cleardoublepage                                                              
\end{center}                                                                  
\end{titlepage}                                                               

\noindent
 C.~Adloff$^{35}$,                
 S.~Aid$^{13}$,                   
 M.~Anderson$^{23}$,              
 V.~Andreev$^{26}$,               
 B.~Andrieu$^{29}$,               
 V.~Arkadov$^{36}$,               
 C.~Arndt$^{11}$,                 
 I.~Ayyaz$^{30}$,                 
 A.~Babaev$^{25}$,                
 J.~B\"ahr$^{36}$
% ...

\noindent
% ...
$ ^2$ III. Physikalisches Institut der RWTH, Aachen, Germany$^ a$ \\
% ...
$ ^{11}$ DESY, Hamburg, Germany$^a$ \\
% ...
$ ^{13}$ II. Institut f\"ur Experimentalphysik, Universit\"at Hamburg,
                                     Hamburg, Germany$^ a$  \\
% ...
$ ^{23}$ Physics Department, University of Manchester,
                                          Manchester, UK$^ b$\\
% ...
$ ^{25}$ Institute for Theoretical and Experimental Physics,
                                                 Moscow, Russia \\
 $ ^{26}$ Lebedev Physical Institute, Moscow, Russia$^ f$ \\
% ...
$ ^{29}$ LPNHE, Ecole Polytechnique, IN2P3-CNRS,
                             Palaiseau, France \\
 $ ^{30}$ LPNHE, Universit\'{e}s Paris VI and VII, IN2P3-CNRS,
                              Paris, France \\
% ...
$ ^{35}$ Fachbereich Physik, Bergische Universit\"at Gesamthochschule
               Wuppertal, Wuppertal, Germany$^ a$ \\
 $ ^{36}$ DESY, Institut f\"ur Hochenergiephysik,
                              Zeuthen, Germany$^ a$\\
% ...
$ ^a$ Supported by the Bundesministerium f\"ur Bildung, Wissenschaft,
        Forschung und Technologie, FRG,
        under contract numbers 6AC17P, 6AC47P, 6DO57I, 6HH17P, 6HH27I,
        6HD17I, 6HD27I, 6KI17P, 6MP17I, and 6WT87P \\
 $ ^b$ Supported by the UK Particle Physics and Astronomy Research
       Council, and formerly by the UK Science and Engineering Research
       Council \\
% ...
$ ^f$ Supported by the Deutsche Forschungsgemeinschaft \\
% ...

\newpage                                                                      
\section{Introduction}                                                        
The measurement of the inclusive deep inelastic lepton-proton                 
scattering (DIS) 
cross section has been of great importance for                      
the understanding of quark and 
gluon substructure of the proton~\cite{Joel}.      
Recently published measurements of the structure function $F_2$ at HERA
based on analyses of the  data collected in 1994
cover the range in squared four-momentum transfer $Q^2$, Bjorken-$x$ and
inelasticity $y$ corresponding to      
$1.5 \le Q^2 \le 5000 $~GeV$^2$, 
$3\cdot 10^{-5} \le x \le 0.32$ and roughly 
$y>0.01$~\cite{F2-h1,F2-zeus,FL-h1}.
These data have shown that $F_2$ continues to rise strongly with decreasing
$x$ for the lowest  $Q^2$ values reached.
Furthermore, it has turned out that 
 the data can be successfully described by perturbative QCD 
in the  measured kinematic range, using the 
leading twist Next to Leading Order QCD evolution
equations, which have subsequently been  used to extract the gluon density 
in the proton at low $x$.

The measured behaviour of $F_2$ at low $Q^2$ is 
 contrary to expectations
 based on  Regge phenomenology~\cite{dola}, which anticipate a much
slower rise with decreasing $x$. At low $x$ the center of  mass 
energy $W$ of the $\gamma^* p$ system is approximately 
$W  \simeq \sqrt{Q^2/x}$.
 Since  $F_2$ at low $x$ is directly 
 proportional  to  the total cross section of the virtual 
photon-proton              interaction $\sigma^{tot}_{\gamma^* p} $,
 the rise of $F_2$ with decreasing $x$ is reflected in a strong 
increase of  $\sigma^{tot}_{\gamma^* p} $ with $W$. The rise can be 
quantified by  parameterizing the data in the form
$F_2 \sim W^{2\lambda}$. For low-$x$ data with $Q^2 \ge 1.5 $ GeV$^2$
and $40< W < 200$ GeV,
values of $\lambda$ in the range of 0.2-0.4 have been measured~\cite{F2-h1}.
This can be contrasted with the increase of the 
real photoproduction ($Q^2 = 0$) 
cross section                         
                              in the same range of $W$~\cite{H1STOT,zeustot}
which gives 
 a value of  $\lambda \simeq 0.08$, 
 similar to that observed for  
hadronic interactions. These processes are dominantly of non-perturbative 
nature and  the total cross section behaviour with $W$ is 
well described by  
Regge phenomenology
 which assumes that the interaction 
dynamics is driven by the so-called soft 
pomeron.
 The $F_2$ data measured in the region $Q^2 \ge 1.5 $ GeV$^2$ 
indicate that the photoproduction limit has not yet been                     
reached. It is therefore of interest to study data at 
still lower $Q^2$ values,
and to experimentally establish and probe the transition  between 
the region of perturbative QCD (DIS) and Regge phenomenology 
(photoproduction),
 a topic
of much theoretical debate at present
(see e.g. \cite {badelek1,levy,ckmt,indurain2}).

The data presented in this  paper extend the $Q^2$ region 
in a continuous way down to lower  values.
 This extension was achieved with the H1 detector
 at HERA  by 
upgrading the detector components  in the backward 
(electron\footnote{HERA operated with $e^-p$ collisions in 1992, 1993
and the start of 1994, and $e^+p$ collisions for the major part of
1994 and all of 1995. In this paper the incident and scattered 
lepton are always referred to as ``electron".} beam direction) 
region. A new calorimeter and a new tracking chamber
with increased acceptance for small scattering
angles were installed in the winter shutdown 1994/1995, and they were
commissioned during the data taking period of 1995. 
In 1995 the incident electron energy was $E_e = 27.5$~GeV and the
proton energy was $E_p=820$~GeV, leading to a total center of mass
energy of the collision of $\sqrt{s}= 300$ GeV.
The data used for this analysis result from a short data taking period where 
the $ep$ collision vertex was shifted by 70 cm in the proton-beam direction
with respect to the nominal vertex position to increase the acceptance 
for the electrons scattered through small angles.
As a result 
the data have 
an acceptance in $Q^2$ down to $Q^2 \sim 0.3 $ GeV$^2$.
In this paper we present data for the kinematic range $0.35 \le Q^2\le 3.5$
GeV$^2$, $x\ge 6\cdot 10^{-6}$ and 
$0.03\le y\le 0.75$, and compare the results with other 
experiments and  with Regge  and QCD inspired phenomenological models.

\section{The H1 Detector}                                       
The H1 detector~\cite{h1detec} is a nearly hermetic multi-purpose apparatus
built to investigate the inelastic high-energy interactions
    of electrons
and protons at HERA. The structure function measurement relies
essentially on the
 tracking chamber system, the backward calorimeter and
the liquid argon (LAr) calorimeter which are described  briefly here.
 
The tracking system includes the central tracking chambers,
the forward tracker modules and a backward drift chamber.
These chambers are placed around the beam pipe at $z$ positions
between --1.5 and 2.5 m. The $+z$-axis is in the 
proton beam direction. A superconducting solenoid
 surrounding both the tracking system and the (LAr)
calorimeter provides a uniform magnetic field of 1.15~T.
 
The central jet chamber (CJC) consists of two concentric drift chambers
covering a polar angle range from  $15^{\rm{o}}$ to $165^{\rm o}$. 
Polar angles are defined with respect to the $+z$ direction.
Tracks
crossing the CJC are measured with a transverse momentum resolution of
${\delta p_T}/{p_T}< 0.01 \cdot p_T$ (GeV).
The CJC is supplemented  by two
cylindrical drift chambers at radii of  18 and 47 cm, respectively,
         to improve the 
determination of  the $z$ coordinate of
the tracks.  A proportional chamber is
attached to each of the $z$ drift chambers for triggering. 
 
A tracking chamber system made of three identical modules 
measures charged particles
 emitted in the forward direction ($7^{\rm o}$ to $20^{\rm o}$). The forward
tracker  is used to determine  the event vertex
for the events which leave no track in the CJC.
This enables vertex reconstruction of events with larger $x$ than can be
achieved with the CJC alone.

In the backward region there is
 an eight layer  drift  chamber (BDC)~\cite{upgrade}
 which has a  polar angle acceptance
of $151^{\rm o}$ to $177.5^{\rm o}$ for collisions at 
the nominal vertex position. 
It replaces the backward multiwire proportional chamber used in 
previous data taking periods.
For data where the  vertex is shifted by 70 cm in the proton direction 
the 
acceptance at large angles increases to a maximum value of $178.3^{\rm o}$. The
BDC provides track segments for charged particles entering the 
backward calorimeter. These are used 
to identify electrons and, together with the 
event vertex reconstructed from tracks in the 
forward and central tracker,
to
measure their polar angle $\theta_e$.
 The  resolution for reconstructed BDC hits
is about 0.5 mm in the radial direction, and 2.5 mm in the 
azimuthal direction.
 
In the backward region a lead/scintillating fiber calorimeter 
(SPACAL)~\cite{spacal} was 
installed, replacing the previous lead/scintillator 
electromagnetic calorimeter (BEMC).
The new calorimeter has both  electromagnetic and hadronic sections.
The angular acceptance of the SPACAL is $153^{\rm o}< \theta<
 177.8^{\rm o}$ for 
collisions at the nominal vertex, increasing to $178.5^{\rm o}$ for 
collisions where the interaction vertex is shifted by  70 cm 
in the proton direction.
The resolution in electron energy is determined using the present data and 
a value of  $7.5\%/\sqrt{E(\mbox{GeV}})\oplus 2.5\%$ is obtained 
for the electromagnetic calorimeter.
The absolute energy scale uncertainty is determined to be 1\% at 27.5 GeV 
increasing  to
3\% at 7 GeV, which is the lowest electron energy used in this analysis.
The high  granularity (1192 cells) results in a
spatial resolution of about 4 mm. 
Around the beam pipe the electromagnetic section 
of the SPACAL has a veto layer (inner and outer 
radius 5.7 cm and 6.5 cm respectively), 
which is used to detect electrons for which the
shower is not fully contained laterally.
The hadronic section has 128 cells.
The   hadronic energy scale uncertainty of the  measurement
in the SPACAL is presently
about 7\%.
Both calorimeter parts have a time resolution better than 1 ns which 
facilitates the  reduction of
  proton beam induced background from beam-wall
and beam-gas interactions which occur upstream of the detector.

To summarize, compared with the detector used up to the end of  1994, 
 the upgrade of the detector components in the backward region
 has resulted in an increased acceptance 
of electrons scattered through small angles 
(large $\theta_e$), in improved granularity 
and in improved resolution. When taken
together with the hadronic energy measurement
these features have  greatly 
facilitated the recognition and removal of background
when selecting DIS events, and have 
extended the available kinematic range of the 
measurement.

Hadronic final state energies are
further measured in the 
LAr calorimeter~\cite{LARC} which covers an  angular
region between $4^{\rm o}$ 
and $154^{\rm o}$. The calorimeter consists of
an
electromagnetic 
       section with lead absorber plates and a hadronic section with
stainless steel absorber plates. 
Both sections are highly segmented in the transverse and longitudinal
directions with about 44000 cells in total.
The electromagnetic part has a depth of between
20 and 30 radiation lengths. The total depth of both calorimeters
varies between 4.5 and 8 interaction lengths. 
The hadronic energy uncertainty of the LAr calorimeter is determined to 
be 4\%.
 
The luminosity is determined from the measured cross section of
      the Bethe Heitler (BH)
reaction $ep \rightarrow ep\gamma$. The final state electron and photon
are  detected in calorimeters
(electron and photon ``taggers") close to the beam pipe but at  large
distances from the main detector (at $z=-33$ m and $z=-103$ m).
For the final value of the luminosity only 
the  hard photon bremsstrahlung data are used.
The precision of the luminosity determination
for these data, where the interaction  vertex is shifted, is
           3\%~\cite{lumipap}.
 It results from the 
error on the luminosity measurement and a correction for 
proton ``satellite" bunches in HERA which 
lead to collisions about 70 cm displaced with respect to the main part
of the proton bunch.

\section{Kinematics}

The kinematic variables
               of the inclusive scattering process $ep \rightarrow eX$
can be reconstructed in different ways using measured quantities from
the hadronic final state and from the scattered electron.
The choice of the reconstruction method for $Q^2$ and $y$
determines the size of systematic errors, acceptance and radiative
corrections. 
The methods used in the analysis of the 1995 data are: i) the
 ``electron  method"
(E), which uses only the event vertex and  the reconstructed scattered
electron, and
 which has the best resolutions in $x$ and
$Q^2$ at large $y$; ii) the  ``$\Sigma$ method"
$(\Sigma)$~\cite{bassler}, which uses the electron and hadronic
final states measurements,
 which
is less sensitive to radiative corrections, 
and which can be used  from very low to large
$y$ values. Both calorimetric and track information are used to calculate
the kinematics with the $\Sigma$ method~\cite{F2-h1}.
The application of different methods is       an important
cross check of the results.

The basic formulae for $Q^2$ and $y$
 in the E method are:
\begin{equation}
  y_e   =1-\frac{E'_e}{E_e} \sin^{2}\frac {\theta_e} {2}
   \hspace*{2cm}
   Q^2_e = 4E'_eE_e\cos^2\frac{\theta_e}{2}
= \frac{E^{'2}_e \sin^2{\theta_e}}{ 1-y_e}, \label{kinematics1}
\end{equation}
where $E'_e$ and $\theta_e$
are the energy and polar angle of the scattered electron.
The formulae for the $\Sigma$ method
are:
\begin{equation}
   y_{\Sigma} = \frac{\Sigma}{ \Sigma + E'_e(1-\cos{\theta_e})}
   \hspace*{2cm}
   Q^2_{\Sigma} = \frac{E^{'2}_e \sin^2{\theta_e}}{ 1-y_{\Sigma}},
\end{equation}
with
\begin{equation}
   \Sigma=\sum_h{(E_h-P_{z,h})}.
\end{equation}
Here   $E_h$ and $P_{z,h}$ are the energy and longitudinal momentum component
of a particle $h$, the summation is  over all hadronic final
state particles, and  masses are neglected.
The denominator of $y_{\Sigma}$ is equal to twice 
the energy of the true incident beam energy,
which differs from the nominal beam energy if one or more real photons
are (mostly collinearly) emitted by the incident electron and not detected. 
The variable $x$ is calculated in both methods as
$x=Q^2/ys$.
The methods  used are the same as those in  the analysis of the 1994 
data which is reported in~\cite{F2-h1}.
%

\section{Monte Carlo Programs and their Implementation} 
Acceptance corrections and background
contributions  are studied with the data and
Monte Carlo simulations.
Monte Carlo DIS events corresponding to twice
the luminosity of the data were generated using             
DJANGO~\cite{django}. This program is based on                          
HERACLES~\cite{heracles} for the electroweak interaction                       
and on  LEPTO~\cite{lepto} to simulate the hadronic                            
final state. HERACLES includes complete 
first order radiative corrections,              
the simulation of real bremsstrahlung photons,
 and the longitudinal             
structure function. 
LEPTO uses the colour dipole model (CDM) as implemented in
ARIADNE~\cite{cdm} which is in good agreement with data on the energy
flow and other characteristics of the final
state as measured by H1~\cite{h1flow} and ZEUS~\cite{zeflow}.
Alternatively, first order QCD matrix elements with additional 
parton showers can be used for the final state QCD radiation.
Hadronization is performed using string fragmentation~\cite{jetset}.
This model does not contain events with large rapidity gaps~\cite{diff},
tentatively interpreted as diffractive events.
Such events can be generated with the 
Monte Carlo programs RAPGAP\cite{rapgap} and DIFFVM~\cite{diffvm}.
The latter generates the diffractive  exclusive channels 
$ep\rightarrow
ep\rho^0$ and $ep\rightarrow
ep\phi$. In the present analysis these channels to a large
extent escape selection. Dedicated measurements of vector meson production
cross sections have been made at HERA~\cite{VM-h1,VM-zeus},
 and are used to  normalize the      
Monte Carlo prediction. 
The program RAPGAP generates events with a continuous mass spectrum of 
diffractive final states and the yield has 
been normalized to the rate of events with a large rapidity 
gap  observed in the data.

The acceptance corrections
were performed using in turn 
the GRV~\cite{GRV} and MRSD0$^{\prime}$ \cite{mrsd0}
parton distributions for the  initial structure functions 
in the Monte Carlo calculations.  
An iterative procedure was used to reweight the input 
structure functions of the Monte Carlo programs with the 
measured $F_2$ values in this analysis, 
as described in Section 6. For the figures in this 
paper which show a
comparison of  the detector response of Monte Carlo with data, 
the result of the final iteration of the input structure function was used.

 Photoproduction background was simulated using  the PHOJET~\cite{phojet}
  generator for 
 $\gamma p$ interactions. 
A sample of photoproduction
 events was generated which contained all  classes of events 
(soft hadronic collisions, hard scattering
processes  and heavy flavour  production), corresponding to three
times the luminosity of the data.
PHOJET was used to generate 
 events  with $Q^2<0.1 $ GeV$^2$, and DJANGO for events
with $Q^2 > 0.1$ GeV$^2$ for acceptance and background calculations. 
The results reported here were found to be insensitive to the value
of the $Q^2$ boundary between the regions where the two generators were used.

The detector response of the Monte Carlo events was simulated in detail 
 with a program based on       the GEANT program~\cite{GEANT}.      
After this step these events were subjected to the same             
reconstruction and analysis chain as the real data.

\section{Event Selection and 
Calibration}                                       
The  data sample used for this analysis 
corresponds to an integrated luminosity of
 114 nb$^{-1}$.
The trigger used requires that there be   a local
energy deposit (cluster) in the SPACAL calorimeter
with energy greater than about
5 GeV occurring in time with an $ep$ bunch 
crossing.
The trigger efficiency is about 99\% for electrons with 
an energy above 7 GeV.
Losses of about  1\% occur due to the event timing requirements.

Deep inelastic events are selected if they satisfy the following criteria:
\begin{itemize}
\item the most energetic cluster in   the electromagnetic section of the 
SPACAL is 
an electron candidate (see below) with a signal
within a  time window  of 
10 ns total width around the expected value for  a
genuine $ep$ collision;
\item a reconstructed event vertex exists, 
as determined using the central or forward 
trackers,
 within 30 cm of the average event vertex position along the 
beam ($z$) direction;
\item for the electron method  
 $\Sigma(E_h-P_{z,h})$ is required to be 
larger than 35 
GeV, where  $E_h$ and $P_{z,h}$ are the energy and longitudinal 
momentum 
of  a particle. The sum is over all energy deposits measured with the 
calorimeters.
\end{itemize}

For the electron candidate the following is required:

\begin{itemize}
\item the energy of the cluster must be larger than 7 GeV;
\item the radius of the cluster\footnote{The cluster radius is defined
as $\sum_i E_i*d_i / \sum_i E_i$, 
where the sum is over all cluster cells; 
$E_i$ is the energy of cell $i$, and
$d_i$ is the distance from the cluster center of gravity to the center of
the cell $i$.}
  must be smaller than
 3.5 cm;
\item there must be a  track segment in the BDC matched to the 
cluster in the SPACAL within $2\,$cm along the radial direction 
and within 2.5 cm along the azimuthal direction;
the segment closest to the center of the cluster is used for the calculation of
$\theta_e$;
\item the 
radial distance from the beamline to the point at which 
 the track associated with the cluster intersects the surface of the 
SPACAL must be
larger than 8.7 cm, corresponding roughly to  an acceptance of 
$\theta_e < 178^{\rm o}$;
\item less than 1 GeV of energy should be  deposited in the veto layer
of the SPACAL 
to avoid having too much loss of energy near the beam pipe;
\item the energy measured  behind the 
electron cluster in the hadronic part of the 
SPACAL, within a radius of 17.5 cm of the projected
electromagnetic shower center, should be less than 0.5 GeV.
\end{itemize}


The selection cuts were designed to have a high efficiency for 
detecting DIS events. For a large part of the kinematic region studied 
the total efficiency is better than 90\%.

The main non-$ep$ backgrounds in the event sample selected are due
to 
proton beam interactions with residual gas and beam line elements              
upstream of the H1 detector.                                                   
An  efficient reduction of the  background is                         
provided by the minimum electron energy and the vertex          
requirements discussed above.  The residual 
non-$ep$ background                         
was  estimated by visual inspection
 to be less than 2\% of the total number of               
 events at the highest $y$, and negligible elsewhere.

\begin{figure}[t]    \unitlength 1mm
\begin{center}
\begin{picture}(160,80)
\begin{sideways} \put(59,62){ {\bf Events}} \end{sideways}  
\put(-60,-3){
\begin{picture}(0,0) \put(62,71){({\large \bf a})} \end{picture}
\begin{picture}(0,0) \put(137,71){({\large \bf b})} \end{picture}
\begin{picture}(0,0) \put(42,5){{\large \boldmath
 $E_{tagger}$ \bf (GeV)}} \end{picture}
\begin{picture}(0,0) \put(122,5){{\large \boldmath $E'_e$ \bf 
(GeV)}} \end{picture}
\epsfig{file=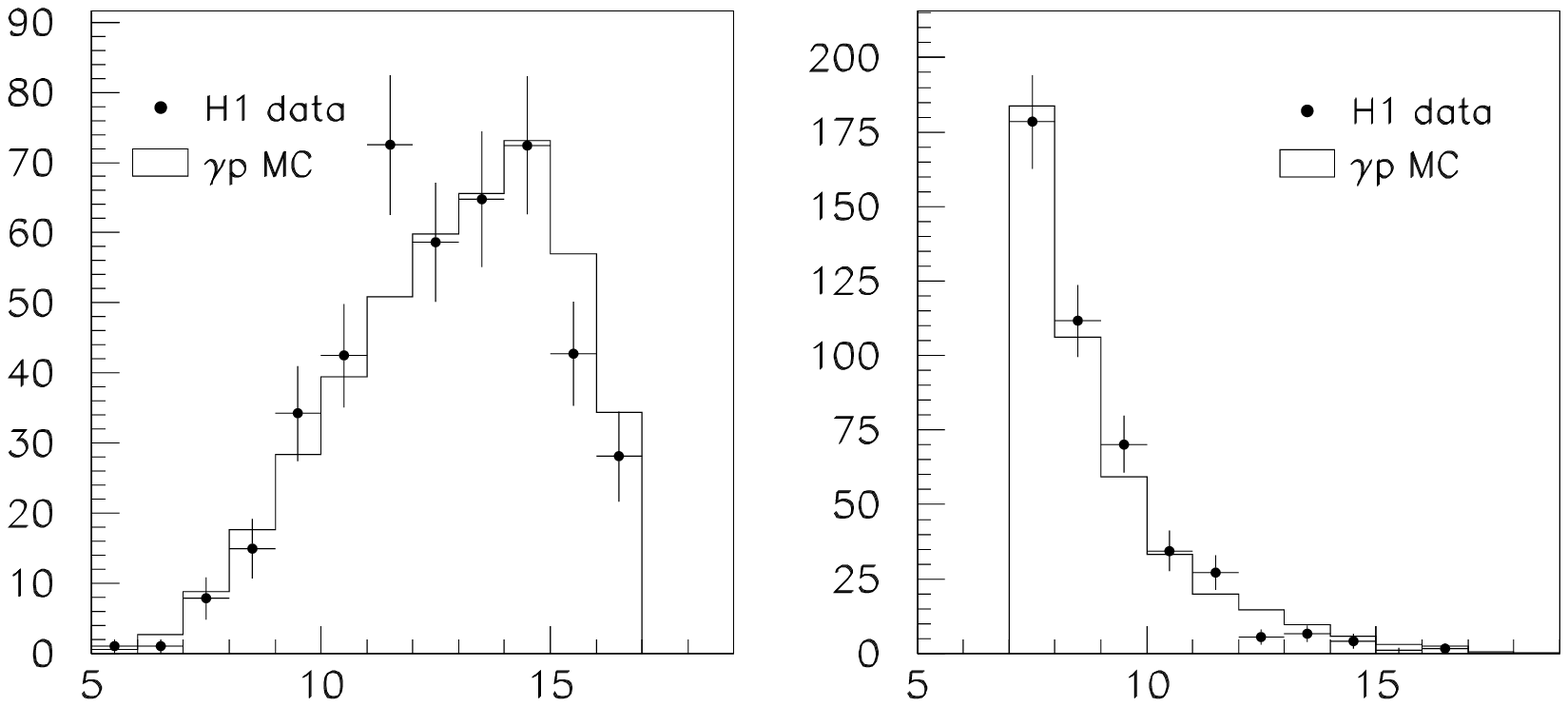,width=13.7cm,bbllx=60pt,bblly=400pt,bburx=500pt,bbury=600pt}   
}
\end{picture}
\end{center}   
\caption[]{\label{fig2}            
\sl  Experimental (points) and Monte Carlo (solid lines)  
distributions of a) the energy in the electron tagger, and 
 b) the energy of the (fake) electron candidate in the SPACAL, for tagged 
photoproduction events.} 

\end{figure}

The only significant background to DIS from $ep$ interactions                  
is due to photoproduction events ($Q^2 \simeq 0$) in which 
the scattered         
electron escapes the main detector along the beam pipe 
and the electron signal is faked by an energy deposition
associated with the hadronic final state.
For about 10\% of these events the scattered electron is detected 
in the electron 
tagger, and  such events can be 
 identified  as photoproduction                 
background.
The total photoproduction 
 background was estimated from studies using simulated events from the 
Monte Carlo program PHOJET.
The results are presented  in Fig.~\ref{fig2}, where the energy spectrum 
of the downstream electron tagger and the 
energy of the (fake) electron candidate in the SPACAL are shown for tagged 
photoproduction events in the  data and in the Monte Carlo predictions.
The Monte Carlo curves are normalized to the number of tagged events.
The prediction for the remaining 
photoproduction background was subtracted 
statistically 
bin by bin. For each $Q^2$ value, only the lowest $x$ bin 
    has a contamination larger than $5\%$. This contamination never           
exceeds $20\%$.                                                     
 
\begin{figure}[b]    \unitlength 1mm
\begin{center}
\begin{picture}(160,80)
\begin{sideways} \put(59,62){ {\bf Events}} \end{sideways}  
\put(-60,-3){
\begin{picture}(0,0) \put(15,71){({\large \bf a})} \end{picture}
\begin{picture}(0,0) \put(92,71){({\large \bf b})} \end{picture}
\begin{picture}(0,0) \put(47,5){{\large \boldmath $E'_e$ 
\bf (GeV)}} \end{picture}
\begin{picture}(0,0) \put(122,5){{\large \boldmath $\theta_e$ 
\bf (degrees)}} \end{picture}
\epsfig{file=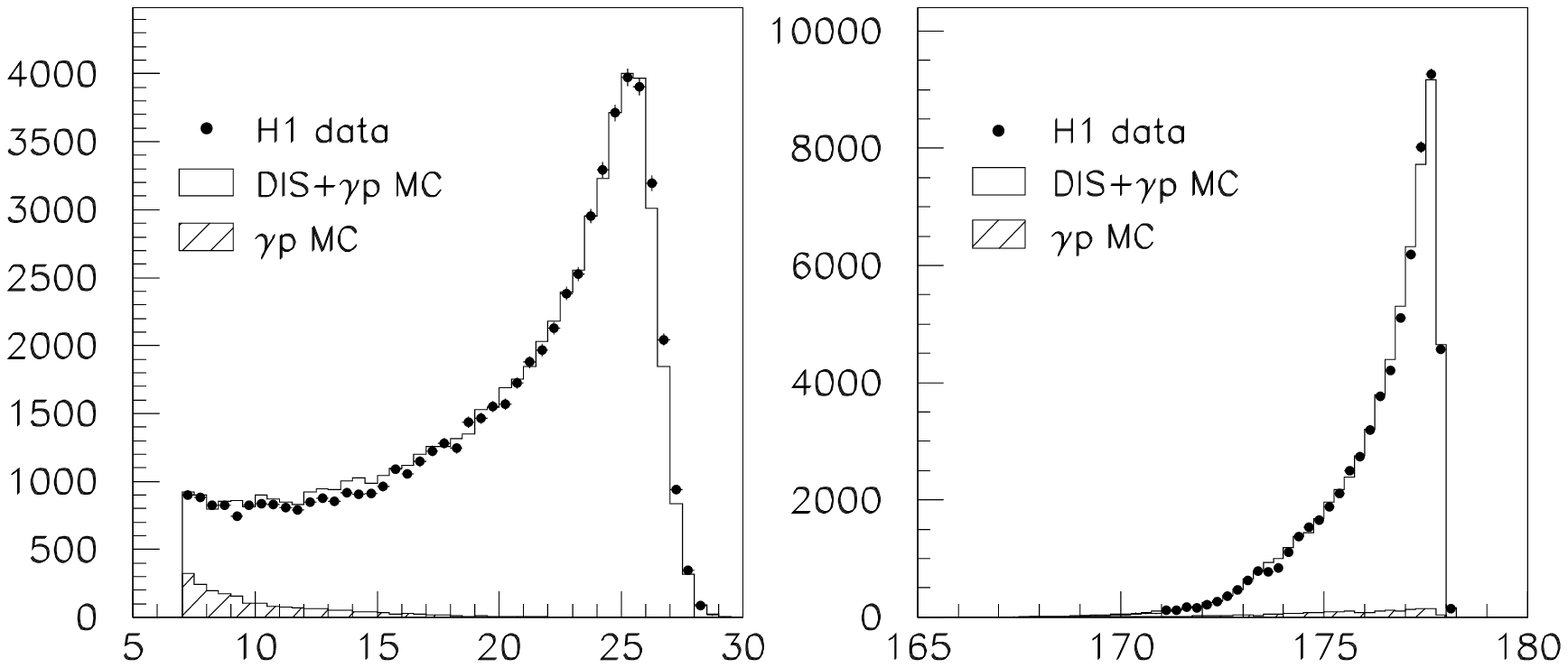,width=13.7cm,bbllx=60pt,bblly=400pt,bburx=500pt,bbury=600pt}   
}
\end{picture}
\end{center}   
\caption[]{\label{fig1}      
\sl  Experimental (points) and Monte Carlo (solid histograms) 
distributions of    a)
  the energy of the scattered electron and 
b) the polar angle  of the scattered 
electron for DIS events. The Monte Carlo 
curves are  the sum of DIS and photoproduction 
events and the photoproduction events alone.}
\end{figure}

\begin{figure}[t]    \unitlength 1mm
\begin{center}
\begin{picture}(160,80)
\begin{sideways} \put(59,62){ {\bf Events}} \end{sideways}  
\put(-60,-3){
\begin{picture}(0,0) \put(62,71){({\large \bf a})} \end{picture}
\begin{picture}(0,0) \put(138,71){({\large \bf b})} \end{picture}
\begin{picture}(0,0) \put(30,5){{\large \boldmath $\Sigma (E-P_z)$ 
\bf (GeV)}} 
\end{picture}
\begin{picture}(0,0) \put(105,5){{\large \boldmath $\Sigma (E-P_z)$ 
\bf (GeV)}} \end{picture}
\epsfig{file=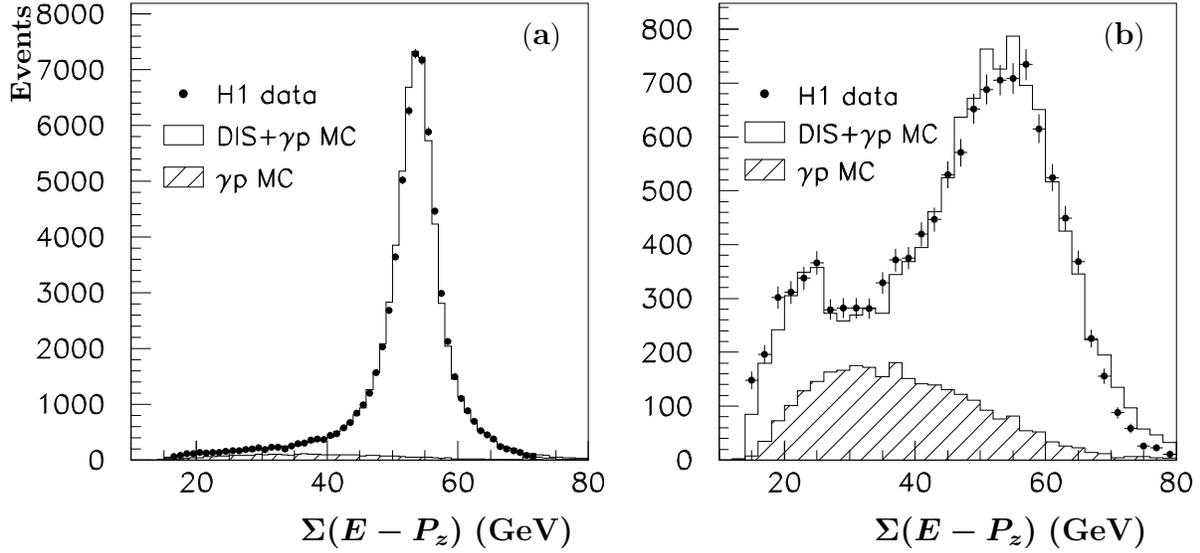,width=13.7cm,bbllx=60pt,bblly=400pt,bburx=500pt,bbury=600pt}   
}
\end{picture}
\end{center}   
\caption[]{\label{fig3}            
\sl  Experimental (points) and Monte Carlo (solid histograms)  
distributions of $\Sigma (E_i-P_{z,i})$ measured in the calorimeter for
a) all DIS event candidates  and b) DIS event candidates with 
$y_e> 0.55$. The Monte Carlo 
curves are the sum of DIS and photoproduction 
events and the photoproduction events alone.
 }
\end{figure}

\begin{figure}[b]    \unitlength 1mm
\begin{center}
\begin{picture}(160,80)
\begin{sideways} \put(63,62){ {\bf Events}} \end{sideways}  
\put(-60,-3){
\begin{picture}(0,0) \put(60,74){({\large \bf a})} \end{picture}
\begin{picture}(0,0) \put(138,74){({\large \bf b})} \end{picture}
\begin{picture}(0,0) \put(50,5){{\Large \boldmath $y_{\Sigma}/y_e$}} 
\end{picture}
\begin{picture}(0,0) \put(125,5){{\Large \boldmath
 $p_{t,h}/p_{t,e}$}} \end{picture}
\epsfig{file=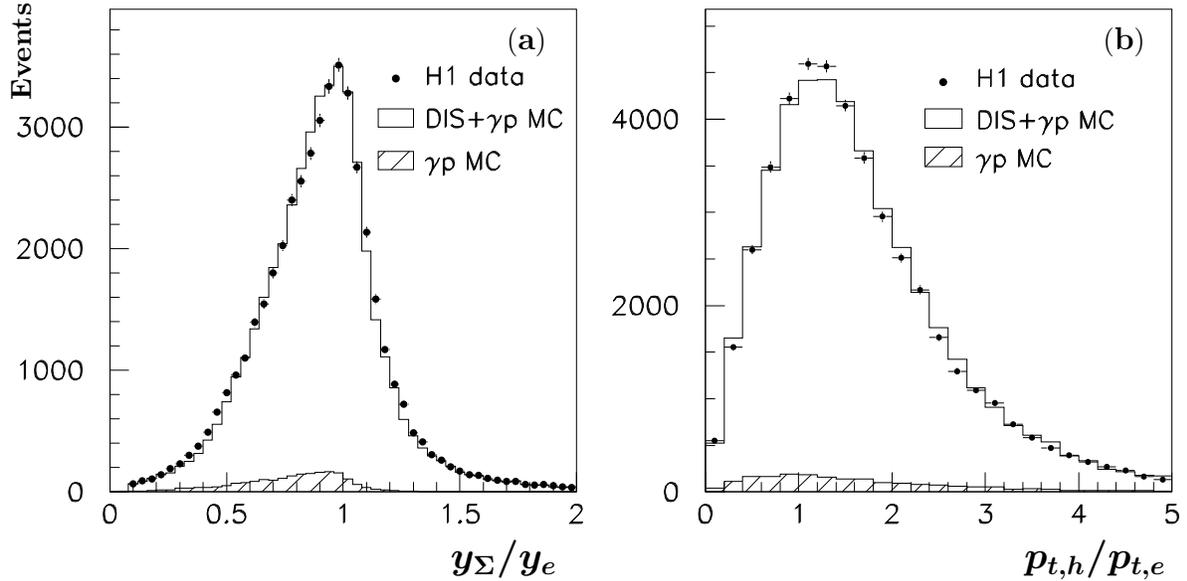,width=14.5cm,bbllx=80pt,bblly=515pt,bburx=530pt,bbury=645pt}   
}
\end{picture}
\end{center}   
\caption[]{\label{fig4}            
\sl  Experimental (points) and Monte Carlo (solid histograms) 
distributions for $y_{\Sigma} > 0.05$ 
of the ratios of a)  the $y$ values measured with the 
$\Sigma$  and E method
 $y_{\Sigma}/y_{e}$, and
b) the transverse momentum of the hadronic system and 
the electron $p_{t,h}/p_{t,e}$.
The Monte Carlo 
curves are  the sum of DIS and photoproduction 
events and the photoproduction events alone.}
\end{figure}

The energy scale for electrons has been determined 
with events at 
low $y$, for which the energy of the scattered electron is very close to
the incident electron energy. The linearity of the 
energy response 
was verified with  QED-Compton events for the energy range used in
this analysis.
The precision of the angular measurement from the BDC 
was estimated using tracks in the central tracker extrapolated into
the BDC region.                    
Fig.~\ref{fig1} shows the energy and angular distribution of electron
candidates in the  sample of selected DIS events, together with Monte Carlo 
predictions for the sum of DIS and  photoproduction 
background in the kinematic range $0.32 < Q^2 < 10$ GeV$^2$.
The structure function of the Monte Carlo calculation is 
reweighted as described in Section 6. For these and  following figures
in this section the Monte Carlo prediction is normalized to the luminosity.
Good agreement is observed between data and Monte Carlo calculations.
Fig.~\ref{fig3} shows the distribution of 
$\Sigma (E_h-P_{z,h})$ for the full data sample and 
for the sample with $y>0.55$. The Monte Carlo calculation describes the data
 well. The  enhancement
 observed around 25 GeV in Fig.~\ref{fig3}b results 
from events with one or more photons 
collinearly emitted by the incident electron.
The level to which the hadronic variables are understood is demonstrated
in Fig.~\ref{fig4}, where the  
distributions of the ratio of the $y$ values measured with the 
$\Sigma$ and E methods,
 $y_{\Sigma}/y_{e}$,  and
the transverse momentum of the hadronic system and the 
 scattered  electron, $p_{t,h}/p_{t,e}$, are shown. Here the data 
are limited to $y_{\Sigma}> 0.05$ to ensure  good quality 
reconstruction of the kinematics with the electron method,
and to the region where the $\Sigma$ method is used for the combined
measurement, namely $0.75< Q^2< 4.2$ GeV$^2$.
In all, the detector response is  
 well understood, allowing a precise extraction of the cross section and
$F_2$.

\section{
Structure Function and  Cross Section 
Measurement}                                       
The  measured $ep$ cross section  in
the HERA kinematic range 
can be expressed in terms of proton 
structure functions or cross sections for virtual photon-proton
interactions as follows
\begin{equation}                                                              
\begin{aligned}
  \frac{d^2\sigma}{dx dQ^2} & =\frac{2\pi\alpha^2}{Q^4x}  
    (2-2y+\frac{y^2}{1+R}) F_2(x,Q^2)  \\
 & = \Gamma [\sigma_T(x,Q^2) +  \epsilon(y)\sigma_L(x,Q^2)]
\equiv \Gamma \sigma^{eff}_{\gamma^*p}(x,y,Q^2).
\label{dsigma}      \\
\end{aligned}                                                           
\end{equation}
  Here $R = F_L/(F_2-F_L)$ where $F_L$ is the longitudinal structure
function, 
 $\alpha$ is the fine structure constant, and $\sigma_L$ and $\sigma_T$
are the cross sections for transverse and longitudinally polarized 
virtual photons. 
The flux factor, $\Gamma$, and the ratio of the 
longitudinal to the transverse flux, $\epsilon$, are taken to be
\begin{equation}
\Gamma = \frac{\alpha (2-2y+y^2)}
{2\pi Q^2x}, \ \ \
 \epsilon(y) = \frac{2(1-y)}{2-2y+y^2}.
\end{equation}
The quantity $\sigma^{eff}_{\gamma^*p}$ is the effective measured 
virtual photon-proton
 cross section for $ep$ collisions in our kinematic range, 
and can be determined from the data without assumptions for $R$.
The total virtual photon-proton cross section
 is here defined\footnote{The exact formula used is 
$\sigma^{tot}_{\gamma^*p}= 
 (4\pi^2\alpha/Q^4)(4M^2x^2+Q^2)/(1-x)\cdot F_2(x,Q^2)$ 
\cite{hand,levyx},  which is approximately
equal to eqn.~\ref{hnd} in the HERA kinematic range; $M$ is the proton mass.}
 as
\begin{equation}
\sigma^{tot}_{\gamma^*p}= \sigma_T(x,Q^2) + \sigma_L(x,Q^2)
\simeq \frac{4\pi^2\alpha}{Q^2}F_2(x,Q^2).
\label{hnd}
\end{equation}
With this definition $\sigma^{tot}_{\gamma^*p}$  depends only on
$Q^2$ and $x$ (or $W$)   
and  the results of different  experiments may easily be compared.

The virtual photon-proton cross section is determined by converting
the measured number of events in a given bin into a bin averaged cross
section using Monte Carlo acceptance calculations.
The data are binned in a grid of $Q^2$ and $x$  for the region $Q^2> 
0.75$ GeV$^2$ as for previous H1 analyses,
 and a grid of $Q^2$ and $y$  for the region 
below 0.75 GeV$^2$, which  optimizes  the access to the  smallest
possible $(Q^2,x)$ values. Bin widths are chosen such that 
the number of events reconstructed in any given  bin
which originate from that bin is larger than 40\% for the E method
and larger than 30\% for the $\Sigma$ method.
 All detector efficiencies are determined from the data                       
utilizing the redundancy of the apparatus. 
Apart from  small               
extra corrections, all efficiencies are correctly                              
reproduced by the Monte Carlo simulation, and therefore the Monte Carlo can
 safely be used to correct for  acceptance and efficiency effects.
The bin averaged cross section is                                             
corrected for first order QED radiative contributions
with the program HERACLES.
The effective virtual photon-proton cross section, 
$\sigma^{eff}_{\gamma^*p}$,
is finally obtained by correcting the bin averaged cross sections for 
each bin to the values at the given bin centers.

 To extract the  structure function $F_2$
from these measurements an assumption  has to be made for the 
longitudinal structure function $F_L$ since it  has  not  yet been measured
in this kinematic region.
In this analysis the model of \cite{BKS} is applied  for 
the calculation of $R$. These values of $R$ are then used
to  determine  the $F_2$ values, as well as  to reweight 
Monte Carlo events. The 
model is based on the photon-gluon fusion process and
has the proper limit for $Q^2 \rightarrow 0$ where $F_L$ should vanish 
$\propto Q^4$. The predictions of this model for
the values of $R$ in our kinematic region vary 
from  0.1 at $Q^2 = 0.35$ GeV$^2$ to  0.3
at $Q^2 = 3.5$ GeV$^2$. The predictions for $R$ 
 at higher $Q^2$ are in agreement with  measurements
from fixed target experiments. 
It should be noted  that this is  a model and future measurements
could reveal  quite different $R$ values. However, 
only the lowest $x$ point at each $Q^2$  is affected 
significantly by 
the assumption made for $R$.  If $R$ 
is taken to be  zero rather than the values obtained using the above model,
the variation in $F_2$ is $5$ to $10\%$ at the highest $y$
(smallest $x$) at a given $Q^2$, and smaller elsewhere.
                                                 
Values of $F_2$
 are derived using  an iterative
procedure.                              
An initial determination of the $F_2$ values uses a
structure function parameterization
in the Monte Carlo simulation as described in Section 3 
and it assumes a value for $R$ as given above.
A new structure function for the full range in $Q^2$ is then calculated 
following the prescription of Badelek and Kwiecinski (BK,~\cite{badelek}), 
where the structure function is 
assumed to be 
the sum of two contributions:  a Vector Meson Dominance (VMD) model term
$F_2^{VMD}$ 
 and a partonic term $F_2^{part}$. The latter becomes 
dominant above $Q^2\sim$ 1 GeV$^2$. 
In this analysis 
the result of a QCD fit similar to that reported in
\cite{F2-h1}  is used for the partonic term.
This fit was made to   structure function data from 
H1~\cite{F2-h1}, and 
in order to constrain the high-$x$ region,  to NMC~\cite{NMC}  and
BCDMS~\cite{bcdms} data. Parton
density parameterizations were defined 
at a starting scale $Q^2_0 = 0.35$ GeV$^2$,
and  data with $Q^2 > 1 $ GeV$^2$ were fitted, yielding
values for $F_2$ denoted as  $F_2^{H1QCD}$ in the following.
The 
newly measured low $Q^2$
data points and  the measurements at  $Q^2=0$ in the $W$ range at 
HERA~\cite{H1STOT,zeustot} were fitted to  the form
\begin{equation}
F_2(x,Q^2)= C_{VM}F_2^{VMD}(x,Q^2) + 
\frac{Q^2}{Q^2+Q^2_{VM}}F_2^{H1QCD}(\overline{x},Q^2+Q^2_{VM}),
\end{equation}
with 
$\overline{x} = (Q^2+Q^2_{VM})/(W^2+Q^2+Q^2_{VM})$.
The fit parameters are the 
meson mass cut-off parameter $Q^2_{VM}$ 
and the normalization
of the vector meson term $C_{VM}$. The latter parameter is not 
part of the BK model  and was introduced to reproduce the 
real photoproduction data  measured at HERA in 
a phenomenological way. 
 Values of 
$Q_{VM}^2= 0.45$ GeV$^2$ and $C_{VM} = 0.77$
are obtained\footnote{Note that this procedure
does not guarantee a 
 consistent separation of $F_2$ into the 
$F_2^{VMD}$ and $F_2^{part}$ contributions,
as prescribed by  the model.}.
It was checked  that 
no further iteration step was needed.
The Monte Carlo curves discussed in Section 4
are reweighted with these $F_2$ values (and the $R$ values discussed above).

{\small
\begin{table}[htbp] 
\centering
\begin{tabular}{|c|c|*{8}{c|}}
\hline
 $Q^2$  &$x$  & $y$ &
 $W$  & $\kappa\sigma^{eff}_{\gamma^*p}$ & $R$ & $F_2$ &
 $\delta_{stat}$
& $\delta_{syst}$
&$\delta_{tot} $\\
(GeV$^2$) & & & (GeV) & & & & (\%) & (\%) & (\%) \\
\hline 
 0.35& 0.0000061&0.640&240.&0.384&0.10&0.397& 5.5&14.5&15.5 \\ 
 0.50& 0.0000086&0.640&240.&0.473&0.13&0.494& 3.6&11.1&11.7 \\ 
 0.65& 0.0000112&0.640&240.&0.539&0.16&0.568& 3.6&10.4&11.0 \\ 
 0.65& 0.0000164&0.440&199.&0.536&0.15&0.547& 2.9& 8.8& 9.2 \\ 
 0.85& 0.0000138&0.682&248.&0.664&0.19&0.713& 5.0&13.0&13.9 \\ 
 0.85& 0.000020&0.470&206.&0.628&0.19&0.646& 2.6& 6.6& 7.1 \\ 
 0.85& 0.000032&0.294&163.&0.615&0.18&0.621& 2.4& 7.2& 7.6 \\ 
 0.85& 0.000050&0.188&130.&0.576&0.18&0.578& 2.8& 8.1& 8.6 \\ 
 0.85& 0.000080&0.118&103.&0.533&0.17&0.534& 3.1&15.4&15.7 \\ 
 1.20& 0.000020&0.664&245.&0.793&0.23&0.857& 3.5&10.5&11.0 \\ 
 1.20& 0.000032&0.415&194.&0.741&0.22&0.759& 2.6& 7.1& 7.6 \\ 
 1.20& 0.000050&0.266&155.&0.709&0.22&0.715& 2.2& 6.9& 7.2 \\ 
 1.20& 0.000080&0.166&122.&0.626&0.21&0.627& 2.1& 8.9& 9.2 \\ 
 1.20& 0.000130&0.102& 96.&0.569&0.21&0.570& 2.4& 5.1& 5.6 \\ 
 1.20& 0.000200&0.066& 77.&0.525&0.21&0.525& 2.4& 5.2& 5.8 \\ 
 1.20& 0.000320&0.042& 61.&0.531&0.21&0.531& 2.4& 7.9& 8.3 \\ 
 1.50& 0.000032&0.519&217.&0.817&0.25&0.855& 3.2& 8.8& 9.4 \\ 
 1.50& 0.000050&0.332&173.&0.771&0.25&0.783& 2.7& 5.9& 6.4 \\ 
 1.50& 0.000080&0.208&137.&0.687&0.24&0.690& 2.4& 6.9& 7.3 \\ 
 1.50& 0.000130&0.128&107.&0.668&0.23&0.669& 2.5& 6.6& 7.1 \\ 
 1.50& 0.000200&0.083& 87.&0.645&0.23&0.645& 2.5& 7.4& 7.8 \\ 
 1.50& 0.000320&0.052& 68.&0.613&0.23&0.613& 2.5& 6.9& 7.4 \\ 
 1.50& 0.000500&0.033& 55.&0.577&0.23&0.577& 2.5& 7.9& 8.3 \\ 
 2.00& 0.000032&0.692&250.&0.945&0.29&1.048& 5.0&12.9&13.9 \\ 
 2.00& 0.000050&0.443&200.&0.908&0.28&0.939& 3.2& 6.0& 6.8 \\ 
 2.00& 0.000080&0.277&158.&0.743&0.27&0.751& 2.9& 5.8& 6.5 \\ 
 2.00& 0.000130&0.170&124.&0.727&0.26&0.729& 2.7& 7.6& 8.0 \\ 
 2.00& 0.000200&0.111&100.&0.716&0.26&0.717& 2.8& 5.8& 6.4 \\ 
 2.00& 0.000320&0.069& 79.&0.727&0.25&0.727& 2.8& 5.9& 6.6 \\ 
 2.00& 0.000500&0.044& 63.&0.639&0.25&0.639& 2.9& 8.3& 8.8 \\ 
 2.50& 0.000050&0.554&224.&0.973&0.30&1.034& 4.3&11.2&12.0 \\ 
 2.50& 0.000080&0.346&177.&0.932&0.29&0.950& 3.3& 5.4& 6.3 \\ 
 2.50& 0.000130&0.213&139.&0.917&0.28&0.922& 2.9& 7.6& 8.1 \\ 
 2.50& 0.000200&0.138&112.&0.840&0.27&0.842& 2.8& 6.8& 7.4 \\ 
 2.50& 0.000320&0.086& 88.&0.702&0.27&0.703& 3.1& 5.3& 6.1 \\ 
 2.50& 0.000500&0.055& 71.&0.649&0.26&0.649& 3.2& 6.8& 7.6 \\ 
 2.50& 0.000800&0.035& 56.&0.590&0.26&0.590& 3.3& 8.2& 8.9 \\ 
 3.50& 0.000080&0.484&209.&1.045&0.32&1.094& 4.1& 8.6& 9.5 \\ 
 3.50& 0.000130&0.298&164.&1.003&0.31&1.018& 3.5& 5.0& 6.1 \\ 
 3.50& 0.000200&0.194&132.&0.869&0.30&0.873& 3.3& 7.2& 7.9 \\ 
 3.50& 0.000320&0.121&105.&0.898&0.29&0.899& 3.2& 8.5& 9.1 \\ 
 3.50& 0.000500&0.077& 84.&0.863&0.28&0.864& 3.4& 6.5& 7.3 \\ 
 3.50& 0.000800&0.048& 66.&0.686&0.28&0.686& 3.6& 7.4& 8.2 \\ 
 3.50& 0.001300&0.030& 52.&0.663&0.27&0.663& 3.6& 7.5& 8.3 \\ 
\hline
\end{tabular}
\caption{\sl  Proton  structure function $F_2(x,Q^2)$ and
effective virtual 
photon-proton cross section $\sigma^{eff}_{\gamma^*p}(W,Q^2)$,
 scaled by the kinematic
factor $\kappa = Q^2/(4\pi^2\alpha)$,  
with 
 statistical ($\delta_{stat}$), 
 systematic ($\delta_{syst}$) and  total ($\delta_{tot}$) 
 fractional  errors. 
The normalization 
uncertainty,  not included in the systematic error, is 3\%. 
The values of $R$ used to calculate $F_2$ are also given.}
\label{tabf2}
\end{table}                                                
}

A list of sources of systematic errors in the  $F_2$ determination 
is given below.                                                         
                                                                               
\begin{itemize}                                                                
\item{Uncertainty of the electron energy scale               
in the SPACAL, varying from 1\% at large electron energies to
3\% at 7 GeV.}
\item{A 4\% scale error for the                                                
hadronic energy in the LAr calorimeter,                                        
the effect of which                                                            
is reduced    due to the joint consideration                                   
of tracks and calorimeter cells for the $\Sigma$ analysis.                     
A  $7$\% scale error was assigned to 
the                                      
 energy of the hadronic final state                                            
measured in the SPACAL. }
\item{A potential shift of up to $0.5$~mrad for the electron polar 
angle.               } 
\item{For the electron identification efficiency 
 the error  was taken 
to be  30\% of the 
 fraction of events lost by the cuts, as given
by the DIS Monte Carlo.}
\item{                                                     
The following contributions to the systematic 
errors from the event selection were included:
trigger and timing veto 0.5\%; BDC efficiency 2\%; 
vertex finding efficiency 2\%.
For the region  $y < 0.05$ the systematic error on the vertex finding
efficiency was increased to 5\%.}
\item{For the radiative corrections an error of 2\% is taken 
everywhere, except for  the highest $y$ point of each $Q^2$ bin 
and for all points with
$Q^2\le 0.65$ GeV$^2$, 
where it  is 
increased to 5\% for the electron method.
This error is                       
due to uncertainties in the hadronic corrections, in the                 
cross section extrapolation                                                    
towards $Q^2 = 0$, in the  higher order corrections and 
the absence of         
soft photon exponentiation in the HERACLES Monte Carlo.
These effects  were
 studied using the  
program HECTOR\cite{hector}.}
\item{The uncertainty due to photoproduction background was assumed            
to be  30\% of the correction applied, i.e.                        
smaller than 6\%. This affects only the highest $y$ bins at 
low              $Q^2$.}
\item{An additional error of 3\% was assigned to the measurements using 
the $\Sigma$ method to allow for the uncertainties in the hadronic final state 
simulation of the Monte Carlo programs. This error was determined by 
comparing the results of different reconstruction methods for the hadronic
final state, using the combined information of calorimeter cells and
tracks, or by using calorimeter cells only.}
\item{The effect of reduced efficiency for detecting diffractive events,
such as the exclusive channel $ep\rightarrow ep\rho^0$, has been
estimated using the  Monte Carlo programs for diffractive processes
discussed 
in Section 4.
 Cross section corrections of up to 6\% are applied
for the points at  the highest $y$ values, and half of the
correction was added to the systematic error.
The effect is largest at the highest $y$ values
where the 
decay products of the meson often escape detection in the CJC and FT, and
hence no event vertex is found.
}
\item{The overall normalization uncertainty is 3\% due to  the 
uncertainty in the luminosity determination}.

\end{itemize}

\begin{figure}[htbp]                                                           
\begin{center}                                                                 
\begin{picture}(0,0) \put(-85,-22){{\boldmath \large $F_2$}} \end{picture}
\begin{picture}(0,0) \put(-85,-75){{\boldmath \large $F_2$}} \end{picture}
\begin{picture}(0,0) \put(-85,-123){{\boldmath \large $F_2$}} \end{picture}
\begin{picture}(0,0) \put(58,-173){{\LARGE \boldmath $x$}} 
\begin{picture}(0,0) \put(-51,-35){{\Large \bf H1-95}}\end{picture}  
\end{picture} 
\epsfig{file=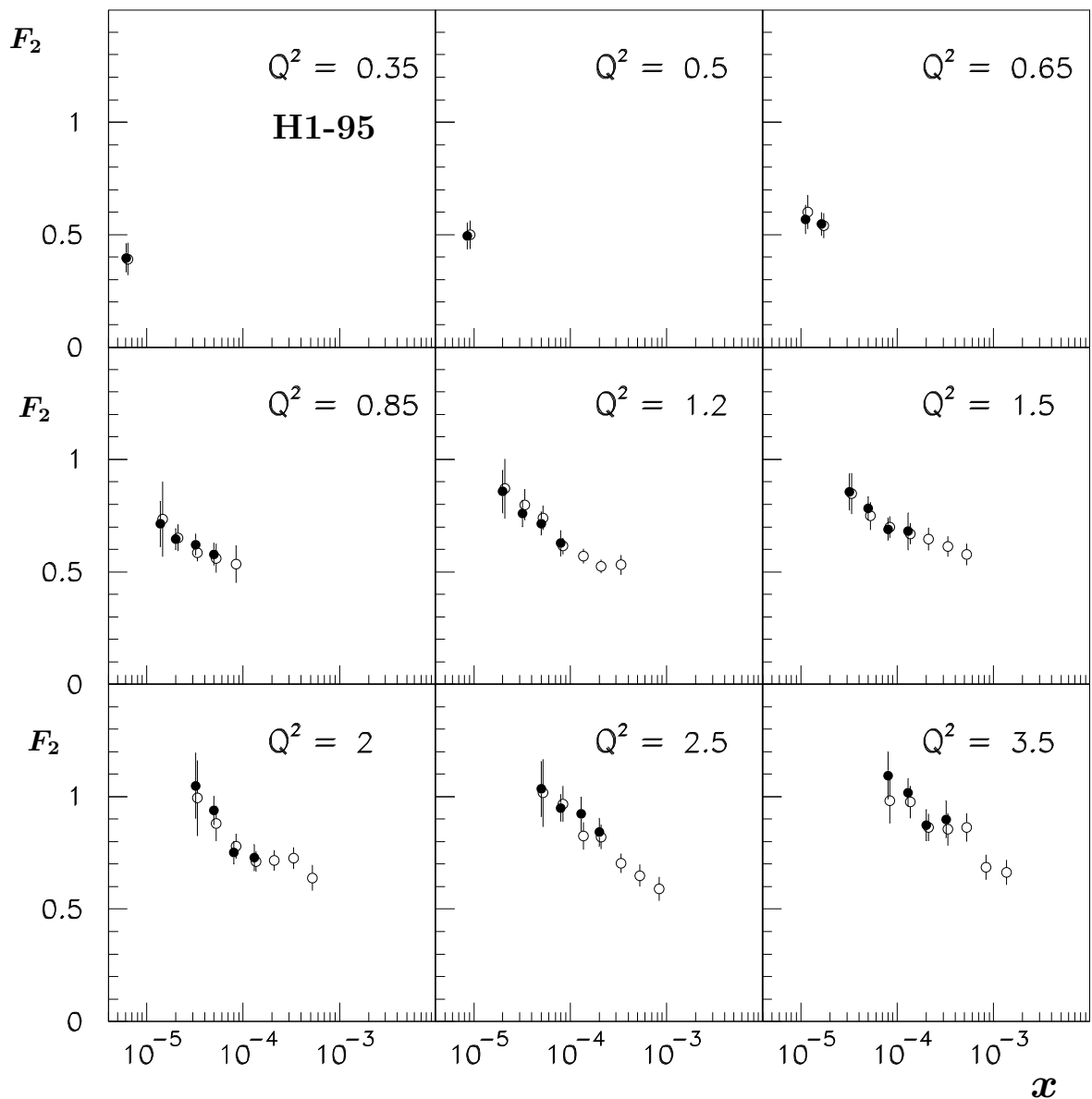,width=15.5cm,%
  bbllx=75pt,bblly=140pt,bburx=550pt,bbury=695pt}                              
\end{center}                                                                   
\caption[]{\label{figb1}                                                      
\sl Comparison of the  proton structure function $F_2(x,Q^2)$ 
as a function of     
    $x$ at various values of    $Q^2$ (in GeV$^2$)
    measured with the E method (full points) and with the $\Sigma$ method      
    (open points).                                                             
    The  errors represent the statistical and systematic errors added      
    in quadrature.
A global normalization uncertainty of 3\% is not included.
}
  \end{figure}

 These systematic uncertainties affect differently the $F_2$                   
measurements made with different methods. In Fig.~\ref{figb1}
the comparison of the measurements made with the E and with the                 
$\Sigma$ method is shown. The statistical and systematic
errors are added in quadrature.
The agreement between the two data sets is very good.

\begin{figure}[htbp]                                                           
\begin{center}                                                                 
\begin{picture}(0,0) \put(-85,-22){{\Large \boldmath $F_2$}} \end{picture}
\begin{picture}(0,0) \put(-85,-75){{\Large \boldmath $F_2$}} \end{picture}
\begin{picture}(0,0) \put(-85,-123){{\Large \boldmath $F_2$}} \end{picture}
\begin{picture}(0,0) \put(58,-173){{\LARGE \boldmath $x$}}\end{picture}  
\begin{picture}(0,0) \put(-52,-35){{\Large \bf H1-95}}\end{picture}  
\epsfig{file=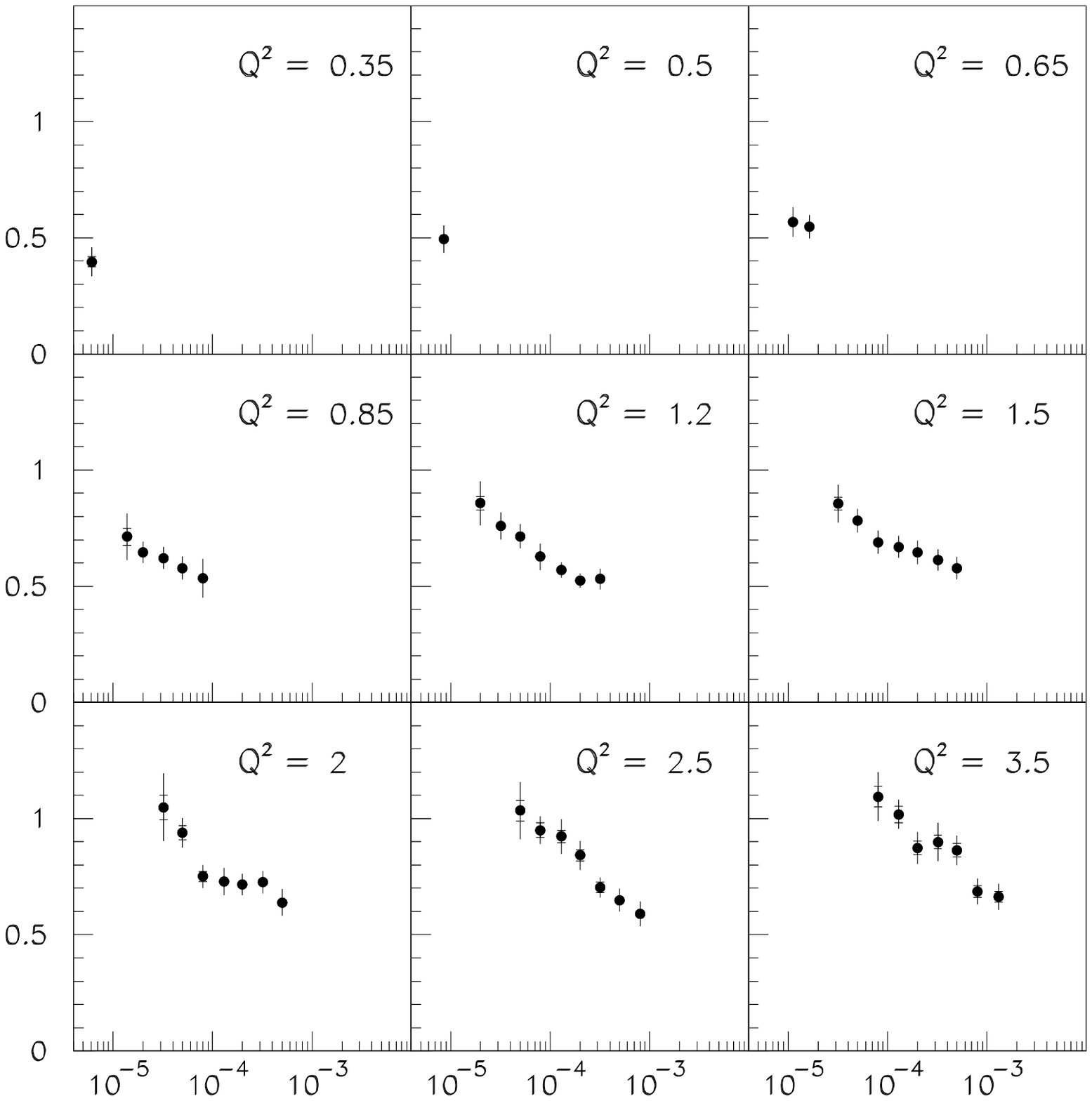,width=15.5cm,%
  bbllx=75pt,bblly=140pt,bburx=550pt,bbury=695pt}                              
\end{center}                                                               
  \caption[]{\label{figb2}                                                    
\sl Measurement of the proton structure function $F_2(x,Q^2)$
    as a function of $x$ at various values of   $Q^2$ (in GeV$^2$). 
The      inner error bars are the statistical errors, the outer
 error  bars           
    represent the statistical and systematic errors added                     
    in quadrature.
A global normalization uncertainty of 3\% is not included.}
\end{figure}

For the final result the values obtained with the E method are 
taken\footnote{This procedure is used 
except for the highest $x$ point  at $Q^2 = 1.5$ GeV$^2$.  
For this point the total error 
calculated with the $\Sigma$ method is almost a factor of two better than that
of  the E method, and
 therefore the $\Sigma$ method is used.}, supplemented with $\Sigma$ points 
 at low $y$
($0.03 < y < 0.12$) where no points from the E method are available.
The result  is shown in Fig.~\ref{figb2} and tabulated in 
Table~\ref{tabf2}. Both  $F_2$ and $\sigma^{eff}_{\gamma^*p}$
 are given. The latter has been multiplied
by $Q^2/(4\pi^2\alpha)$ to demonstrate directly the effect of $R$. 
The table also contains the  statistical, systematic and total 
errors and the
value of $R$ used for the $F_2$ 
calculation.
 A table  delineating the many different correlated
and uncorrelated error contributions is 
available on request to the H1 collaboration.
The  measurements have                                                         
 a typical  systematic error                                      
of 5-10\%.
Compared with the previous H1 analysis~\cite{F2-h1}                            
the $F_2$ measurement has been extended to lower                           
$Q^2$ (from $1.5$~GeV$^2$ to  $0.35$~GeV$^2$),                   
and to lower $x$ (from $3 \cdot 10^{-5}$ to                  
$6 \cdot 10^{-6}$).

\section{Discussion of the Results}

\begin{figure}[htbp]       \unitlength 1mm                               
\begin{center} 
\begin{picture}(0,0) \put(-85,-22){{\Large \boldmath $F_2$}} \end{picture}
\begin{picture}(0,0) \put(-85,-75){{\Large \boldmath $F_2$}} \end{picture}
\begin{picture}(0,0) \put(-85,-123){{\Large \boldmath $F_2$}} \end{picture}
\begin{picture}(0,0) \put(58,-173){{\LARGE \boldmath $x$}}\end{picture}  
\epsfig{file=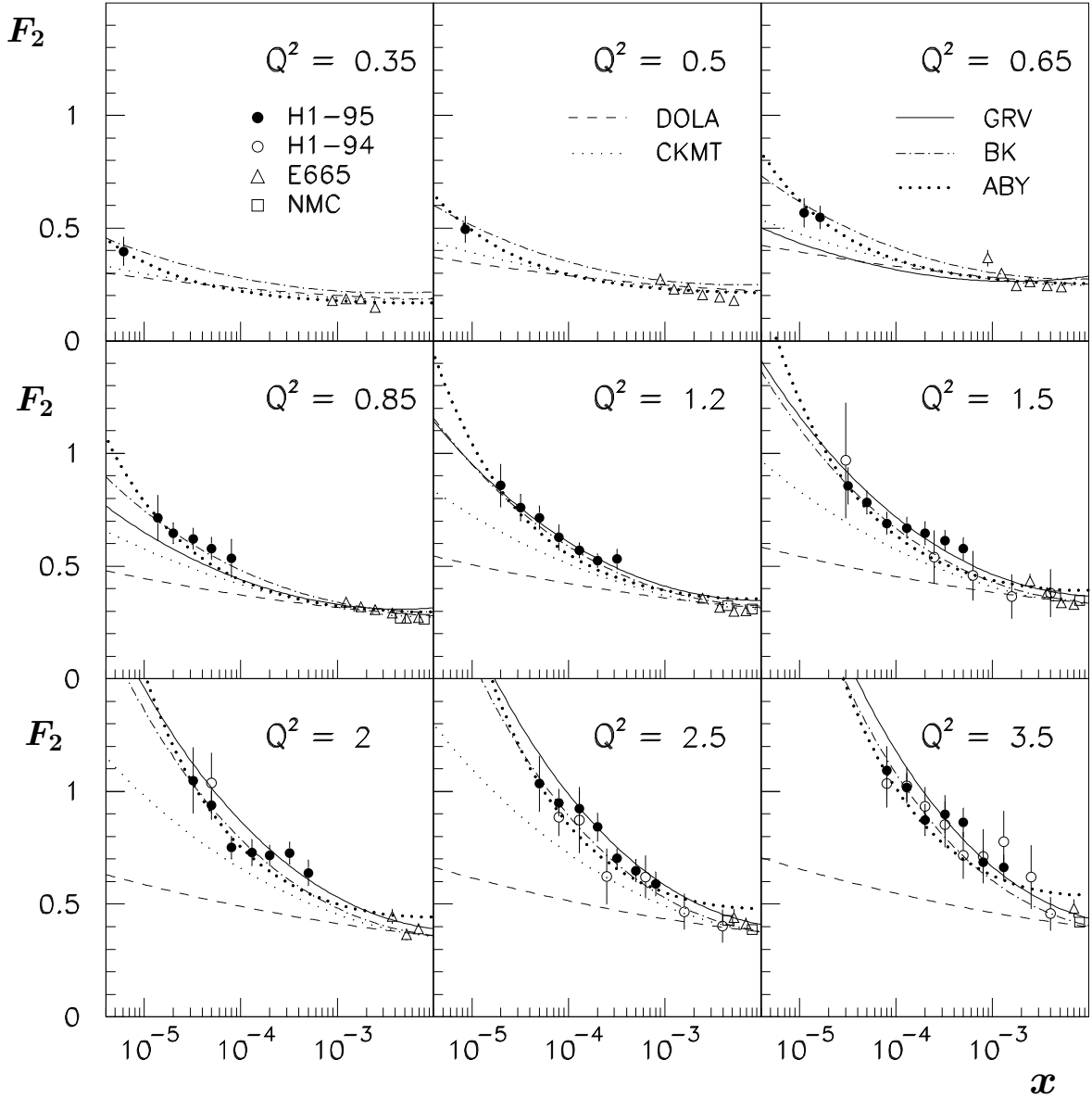,width=15.5cm,%
  bbllx=75pt,bblly=140pt,bburx=550pt,bbury=695pt}                              
\end{center}
\caption[]{\label{figb3}                                                       
\sl Measurement of the proton structure function $F_2(x,Q^2)$
in the low $Q^2$ region by H1 (full points), together with 
previously published results from H1 (open circles), 
E665  (open triangles), NMC (open squares).
The $Q^2$ values are given in GeV$^2$. 
Various  predictions for $F_2$ are compared with the        
data: the model of Donnachie and Landshoff (dashed line), the model
of Capella et al.
(dotted line/small ), the model of Badelek and Kwiecinski (dashed-dotted line),
 the model of Gl\"uck, Reya and Vogt (full line) and the model 
of Adel et al. (dotted line/large).
Global normalization uncertainties 
are not included.
}

\end{figure}

In Fig.~\ref{figb3} the $F_2$ data are compared with previous 
H1 measurements~\cite{F2-h1}, with 
the fixed target measurements of E665~\cite{E665} and 
NMC~\cite{NMC},
 and with the
predictions of models for $F_2$ at low $x$.
 In the region of overlap  
the results are in good                                                       
agreement with our previous measurements and the total error has been 
reduced by a factor of 2 to 3.
The data also show a smooth continuation from the fixed target measurements
towards the low-$x$ region at HERA.
The rise of $F_2$ with decreasing  $x$ is still clearly prominent for
values of 
$Q^2 \ge 2$ GeV$^2$ but becomes less steep for smaller $Q^2$ values. 

\begin{figure}[htbp]                                                           
\begin{center}                                                                 
\epsfig{file=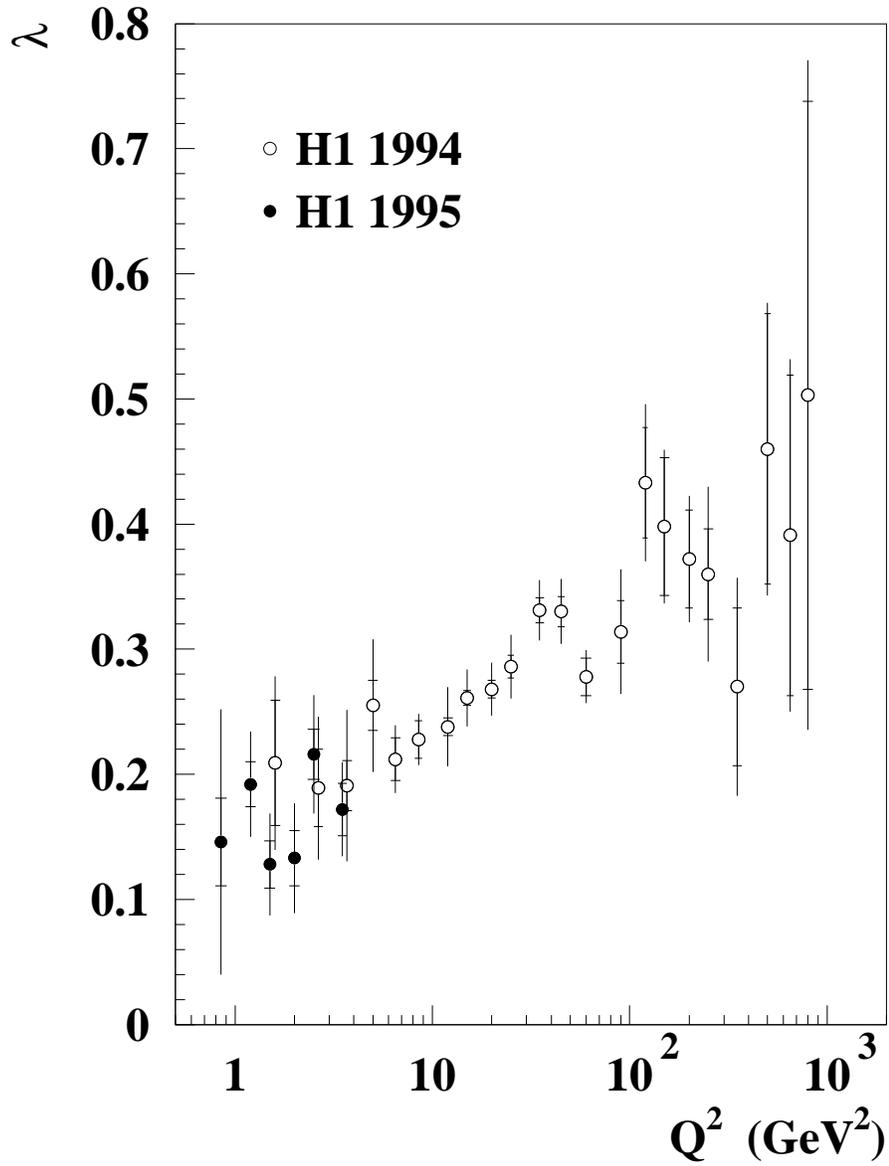,width=16cm,
  bbllx=30pt,bblly=150pt,bburx=560pt,bbury=650pt} 
\end{center}                                                                   
\caption[]{\label{expo}
\sl Variation  of the exponent $\lambda$   from fits
of the form $F_2 \sim x^{-\lambda}$ at fixed $Q^2$
values and $x<$ 0.1. Full symbols are the data from this analysis;
open symbols are the data from  \protect \cite{F2-h1}.
The inner errors are statistical, 
and the full errors represent the statistical 
and systematic errors added in quadrature.  } 
\end{figure}

In Fig.~\ref{expo}, values of 
$\lambda $ are shown from fits of the form 
$F_2 \propto x^{-\lambda}$ 
$(\sim W^{2\lambda})$ at fixed $Q^2$ to the 
present H1 $F_2$ 
data. 
For each $Q^2$ bin with $Q^2\ge 0.85$ GeV$^2$ and $x< 0.1$
the exponent
 $\lambda$ was determined taking into
 account the point to point systematic error correlations. 
Note that this is merely a convenient parameterization to show the
change of the slope with various 
$Q^2$ in a quantitative way. Due to the HERA 
kinematics the $x$ region in which the data are fitted is different for 
the $Q^2$ values used.
The result
 is given in Table \ref{expotab}. 
Fig.~\ref{expo} also includes the H1 measurements of $\lambda$
 reported in~\cite{F2-h1}.
The newly measured values are consistent with the previous low-$Q^2$
measurements, and are in the range   $\sim 0.1-0.2$.

\begin{table}[htbp]
\begin{center}
\begin{tabular}{|c|c|c|c|}
\hline
$Q^2 $ (GeV$^2$) &$\lambda$&$\delta\lambda_{stat}$&$\delta\lambda_{syst}$
\\
\hline
     0.85 &  0.146 &  0.033 &  0.101 \\
     1.2  & 0.192  & 0.018  & 0.038   \\
     1.5  & 0.128  & 0.019  & 0.036   \\
     2.0  & 0.133  & 0.022  & 0.038   \\
     2.5  & 0.216  & 0.020  & 0.043   \\
     3.5  & 0.172  & 0.021  & 0.031   \\
\hline
\end{tabular}
\caption{\sl  The values of the exponent $\lambda$ as a function
of $Q^2$, together with the statistical ($\delta\lambda_{stat}$)
and systematic ($\delta\lambda_{syst}$) contribution to the error.
}
\label{expotab}
\end{center}
\end{table}

\begin{figure}[htbp]                                                           
\begin{center}                                                                 
\begin{sideways} \put(-50,250){ {\LARGE
 \boldmath $\sigma_{\gamma^* p}^{tot} \bf (\mu b)$}} 
\end{sideways}  
\epsfig{file=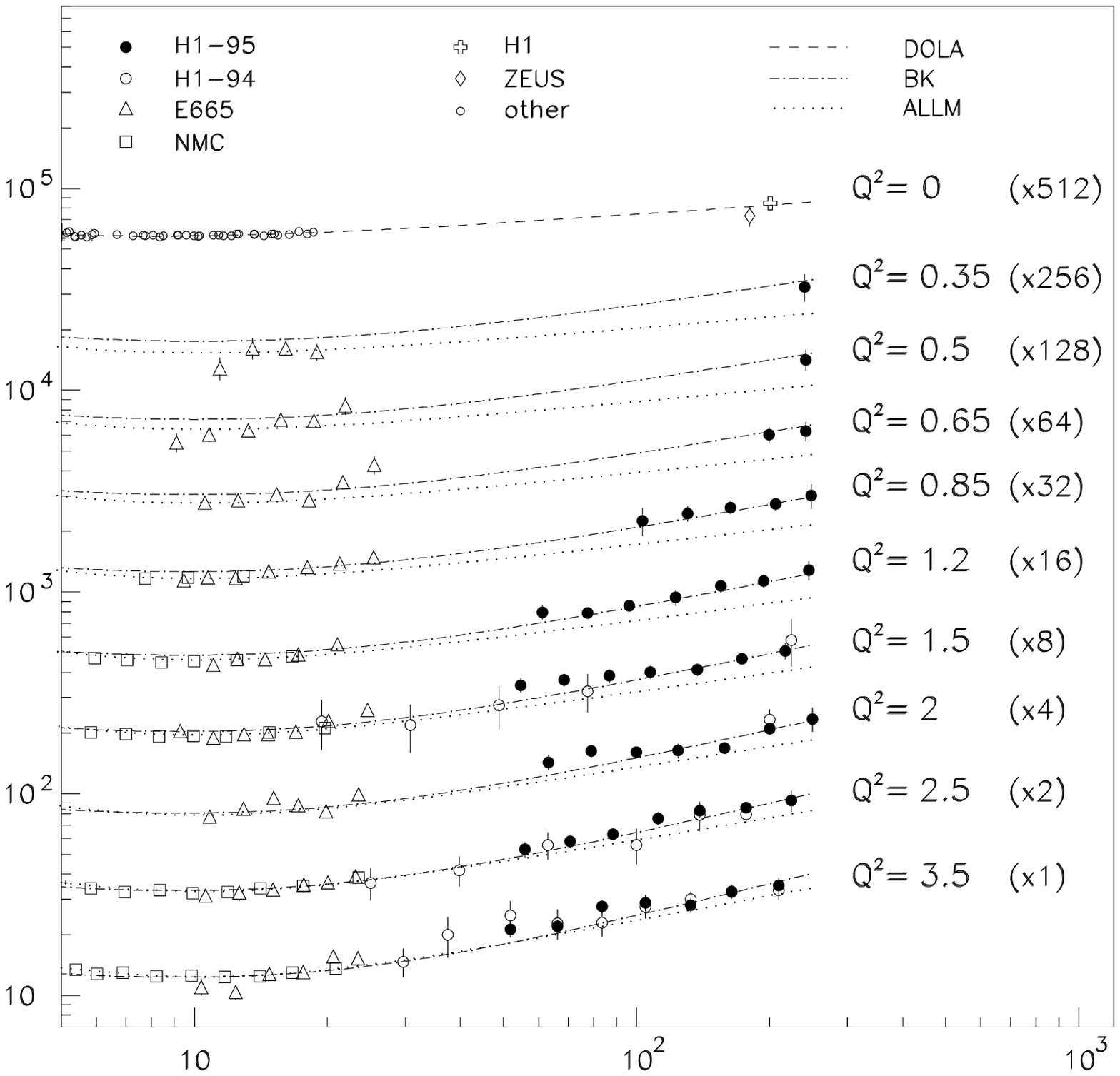,width=15.5cm,bbllx=25pt,bblly=140pt,bburx=550pt,bbury=695pt}                              
\end{center}                                                               
\begin{picture}(0,0) \put(120,15){{\Large \boldmath $W$ \bf 
(GeV)}}\end{picture} 
  \caption[]{\label{fig10}                                                    
\sl  Measurement of the total virtual photon-proton cross section
 $\sigma_{\gamma^*p}^{tot}$
    as a function of $W$ at  various values of  $Q^2$ (in GeV$^2$). 
The cross sections  are multiplied
with the factors indicated in the figure (numbers in brackets).
The  errors               
    represent the statistical and systematic errors added                     
    in quadrature. Full symbols are  data from this analysis;
open symbols are previously published  data from H1 (circles), E665
(triangles) and NMC (squares).
Global normalization uncertainties
are not included.
The curves represent the DOLA (dashed line, only 
shown for $Q^2=0$), ALLM (dotted line)  and BK 
(dashed-dotted) parameterizations.}

\end{figure}

Fig.~\ref{fig10} shows $\sigma^{tot}_{\gamma^*p}$ as a function of the
$\gamma^*p$ center of mass energy $W$ for fixed values
of $Q^2$.
The  H1 data are compared with low energy measurements and
with photoproduction data. The low energy measurements are at slightly
different $Q^2$ values and were propagated to the values indicated
on the figure using the ALLM parameterization~\cite{ALLM} (see below). 

In Fig.~\ref{fig11} 
$\sigma^{eff}_{\gamma^*p}$ is shown as a function of $Q^2$
for $W$ values above 60 GeV. The data are transformed to the
$W$ values given using the ALLM parameterization.
The new results presented in this analysis help to 
close  the gap between measurements of real and virtual
photoproduction and
provide further information on the transition  between 
real and virtual photon interactions.


Several
phenomenological parameterizations
based on models have been proposed to describe the region at low $Q^2$ 
and the transition from photoproduction to DIS,  often using
  ingredients both
from Regge theory at  low $Q^2$,  and from QCD when $Q^2$ is of the
order of 1~GeV$^2$ or larger (for a recent review, see e.g. 
\cite{levy}). 
 A smooth transition from the photoproduction to the
DIS regime is generally assumed, but there is still
uncertainty as to  how this 
transition takes place and 
as to  the underlying dynamics.

Parameterizations motivated by Regge theory
 relate the structure function to
Reggeon and pomeron exchange phenomena which successfully describe
 the slow rise of the total
cross section with the center of mass system energy
in hadron-hadron and real photon-proton interactions.
The model of Donnachie and Landshoff~\cite{dola} (DOLA) 
 assumes a  pomeron 
intercept of $1 + \lambda \sim 1.08$ for the energy dependence of 
the cross section, independent of the virtuality of the photon.
Fig.~\ref{figb3} shows that
the model gives, as expected from the $\lambda$ measurements as 
a function
of $Q^2$, 
  a prediction of  $F_2$ which is 
 too low for the region of $Q^2\ge 0.85$ GeV$^2$, but 
which approaches  
the data for the lowest $Q^2$ values. 
The DOLA prediction 
 can also be interpreted
 as  the  contribution  of  soft 
pomeron exchange to the cross section at non-zero $Q^2$.

In the approach of Capella et al.~\cite{ckmt} (CKMT)
 it is assumed that screening corrections and multi-pomeron
exchange contribute less with increasing $Q^2$, leading to 
a $Q^2$ dependent power $\lambda$.
Hence, the structure 
function $F_2$ predicted by CKMT rises faster with decreasing 
$x$ for increasing 
$Q^2$  than the DOLA 
calculations. 
Furthermore, the model assumes that this prescription accounts for
non-perturbative contributions to $F_2$ at a $Q^2_0$ scale of about 2 GeV$^2$, 
from where, at higher $Q^2$, perturbative QCD
(pQCD) evolution equations  are applied to predict the 
$Q^2$ dependence of $F_2$.
As expected, Fig.~\ref{figb3} shows  that the prediction\footnote{
The CKMT curves shown in Fig.~\ref{figb3} were calculated using the 
parameters given in ~\cite{ckmt} without QCD evolution. Therefore
the prediction is not shown for the highest $Q^2$ value.}
is  systematically above that  from DOLA, but it is 
still below the 
data for all but the lowest values of $Q^2$.

 The parameterization of Abramowicz et al.~\cite{ALLM}
(ALLM) 
 assumes that the total $\gamma^* p$ cross section 
consists of  two contributions
which distinguish  Reggeon and pomeron exchange 
 and the behaviour of the power
$\lambda$  
is assumed to vary with $Q^2$ in a logarithmic way, emulating
pQCD evolution in the high $Q^2$ region.
In Fig.~\ref{fig10} it is shown that the prediction is below  the
data for small $Q^2$ values, but  agrees with the data for 
$Q^2 > 2$ GeV$^2$.

Contrary to these Regge inspired models,
the model of Gl\"uck, Reya and
Vogt~\cite{GRV} (GRV) is defined completely within the parton picture. 
It is  assumed  that all parton distributions at
very low  $Q^2_0= 0.34$ GeV$^2$
 have a valence-like shape, i.e. vanish for $x\rightarrow 0$. 
Assuming further that $F_2$ evolves towards large $Q^2$ values via parton
radiation which is given by 
the leading twist QCD evolution  equations, 
 GRV predict that
   the structure
 function $F_2$ should rise with decreasing  $x$, even for low values 
of  $Q^2\sim 1 $ GeV$^2$.
Fig.~\ref{figb3} shows that the GRV distributions describe the  data well 
for $Q^2 \ge 1$ GeV$^2$, but are systematically lower than the data 
for $Q^2 < 1$ GeV$^2$. At low $Q^2$ values 
 the QCD evolution is over a range of $Q^2$ which is too
 small, and the parton distributions become 
dominated by the valence behaviour at the starting scale $Q^2_0$.

The model of  Badelek and Kwiecinski~\cite{badelek} (BK)
 combines the concepts of Vector Meson Dominance (VMD) 
with dynamical parton  models such as that  of GRV. 
It has per force a smooth transition 
from pQCD to the real photon limit, which
 coincides with the region measured here.
The BK model predicts the 
$F_2$ values very well as shown in Fig.~\ref{figb3}.
In Fig.~\ref{fig10} 
 BK is compared with the
total $\gamma^*p$ cross section measurement
and also  shows  good agreement.

A different  approach to  the low $Q^2$ 
behaviour of $F_2$ in the transition region between photoproduction and DIS
has been presented in~\cite{indurain2} by Adel, 
Barreiro and Yndurain (ABY).
It assumes
that perturbative  QCD  evolution
is   applicable  to the lowest values of $Q^2$.
In addition to the flat or so called 
``soft" behaviour of $F_2$ with decreasing $x$ for  $Q^2\sim 1 $ GeV$^2$,
which is manifest in for example the GRV model, a ``harder'' contribution
 is introduced to 
prevent  $F_2$ decreasing with decreasing $x$ 
for $Q^2$ values below 1 GeV$^2$. In~\cite{indurain2}
the prescription for this 
contribution is  singular $\sim x^{-\lambda_s}$, 
with $\lambda_s = 0.48$ independent of $Q^2$.
Furthermore  the strong coupling constant is assumed to 
become independent of $Q^2$ for values below roughly 1 GeV$^2$,
that is $\alpha_s$
``saturates''. The result of a fit of the ABY
approach to data, reported in  ~\cite{indurain2},
is compared with the new $F_2$ results
presented here  in Fig.~\ref{figb3}. There is 
 good agreement at low $x$, but also 
some possible disagreement in the higher ($x,Q^2$)
region shown in the figure.
Fig.~\ref{figb3} also  shows that in this 
prescription,  at low $Q^2$, the rise of $F_2$ with decreasing $x$
occurs at lower $x$ compared to the other approaches,
namely for  $x <10^{-4}$.

In Fig.~\ref{fig11}, BK and ALLM are compared with the measurements
of $\sigma^{eff}_{\gamma^*p}$.
To ensure consistency, 
for the BK  prediction the value of $R$ has been taken from~\cite{BKS}.
The ALLM prediction as calculated here 
is for $R=0$ and therefore it is not shown for the largest 
$W$. For the other values of $W$ 
the effect of $R$ is a few percent. 
 The measurements suggest that in the
current  ALLM parameterization the 
transition  towards photoproduction behaviour occurs at too large a $Q^2$,
in contrast to the BK approach.
 However  BK predict a photoproduction cross section which is  larger
than the   measurements, as shown in the figure. 
Fig.~\ref{fig11} also shows the 
H1 fit based on BK, as discussed in Section 6.
The photoproduction cross section 
measurements at HERA  were used in the fit, which 
leads to  a stronger  turn-over
to the photoproduction regime compared to the original 
BK  
model.  This  fit gives a good phenomenological description of all data
shown in the figure.

In summary,
it turns out that the region $0.1 < Q^2< 1$ GeV$^2$
spans the kinematic range 
in which Regge or VMD inspired models describe the 
data at  low $Q^2$, and models based on pQCD account well
for the higher $Q^2$ domain. Although in general several of these
models show the correct qualitative behaviour of the data,
none of them  gives at present a  completely 
satisfactory description of the data. Future 
studies will have to show if this can be remedied by 
adjusting the parameterizations whilst simultaneously preserving their internal
consistency. 
The data in the low $(x,Q^2)$  region will help to discriminate 
 between different theoretical approaches to low-$Q^2$ dynamics.

Finally  a  parameterization is given  of H1 data
in the region $ 0.35 \le Q^2 \le 5000$ GeV$^2$ and $y \ge 0.01$.
Starting from the double asymptotic expression for $F_2$ as developed 
in~\cite{ball}, a QCD inspired parameterization with only two parameters
was reported in~\cite{deroeck}.
 Here, $F_2$ is parameterized as
\begin{equation}
F_2 = N_fx^{-\gamma\sqrt{T/\ln(1/x)}} \label{eq1}
\end{equation}
with $T=\ln[\ln((Q^2+Q^2_0)/\Lambda_{eff}^2)/\ln(Q^2_0/\Lambda_{eff}^2)]$
 and $N_f= 5n_f\sqrt{\gamma/\pi}/324$,
where $n_f$ is the number of quark flavours, taken to be 
equal to four. The parameter $\Lambda_{eff}$ resembles the QCD mass scale.
The function $T$ has been slightly changed with respect
to the original proposal~\cite{deroeck} 
so as to achieve a smooth behaviour 
 over the whole $Q^2$ range. The parameter $\gamma$ is defined as
$\sqrt{12/(11-2n_f/3)}$. 
 Fitting this expression to 
the H1 data gives $Q^2_0 = 373 \pm 24$ $ (stat.)\pm 42 $ $(syst.)$ MeV$^2$  and
$\Lambda_{eff} = 247 \pm 11 $ $(stat.)$  $ \pm 19$ $(syst.)$
MeV, with a $\chi^2/ndf = 156/231$
(full errors). 
Athough eqn.~\ref{eq1} does not satisfy the constraint,
 imposed by current conservation, that $F_2$ should 
vanish at $Q^2 = 0$,
it constitutes a compact QCD-inspired parameterization
of all  H1 measurements of $F_2$.


\section{Summary}                                                             
A measurement has been presented of                                           
 the proton structure function $F_2(x,Q^2)$ and virtual photon-proton
cross section $\sigma^{eff}_{\gamma^* p}(x,y,Q^2)$ in 
deep inelastic electron-proton scattering 
at low $Q^2$ with data                    
taken in the 1995 HERA running period.
These are the first measurements  made with the 
upgraded backward calorimeter and drift chamber of the H1 detector.

The measurements presented                                                   
are obtained using two different methods to reconstruct the inclusive         
scattering kinematics, allowing  both a   powerful 
internal cross check of the data  and 
the measurement in  a large kinematic region.
 The data cover the region 
 of  $Q^2$ between 0.35~GeV$^2$ and
3.5~GeV$^2$ and with  Bjorken-$x$ values down to $6\cdot 10^{-6}$.
The measurements
 show a smooth transition from the fixed target high-$x$ data        
to the HERA low-$x$ data.
                                                                             
The 
distinct rise  of the structure function with decreasing                     
$x$ in the low-$x$ region, which is very prominent for  $Q^2\ge
2 $
GeV$^2$,
diminishes  at lower   $Q^2$ values. 
When taken together with the data from 
fixed target experiments, the rise observed for  the smallest 
$Q^2$ values  approaches that  expected in 
Regge and VMD interpretations.

The data have been compared with different models which aim to describe the 
whole $Q^2$ region. Several of these models predict the correct qualitative
behaviour observed in the data but presently do not agree with the data 
throughout the full kinematic range.
The data access  the transition region 
from 
DIS to photoproduction and provide powerful constraints on the development
 of further
 low-$Q^2$ phenomenology.

\vspace*{1.cm}                                                                 
{\bf Acknowledgements}                                                         
\normalsize                                                                    
                                                                               
\noindent                                                                      
 We are very grateful to the HERA machine group whose outstanding              
 efforts made this experiment possible. We acknowledge the support             
 of the DESY technical staff.                                                  
 We appreciate the substantial
 effort of the engineers and                             
 technicians who constructed and maintain the detector. We thank the         
 funding agencies for financial support of this experiment.                    
                       We wish                   to thank the DESY             
 directorate for the support and hospitality extended to the                   
 non-DESY members of the collaboration. 
 We are grateful to  B. Badelek and A. Vogt for providing us with the
 calculations of  $R$ (BK) and $F_2$ (GRV), respectively,
 in the region of our data.

\newpage                                                                       


\newpage                                                                       
\newpage                                                                       
\newpage                                                                       
        
\newpage                                                                       \begin{figure}[htbp]                                                           
\begin{center}                                                                 
\begin{sideways} \put(-50,230){ {\LARGE
 \boldmath $\sigma_{\gamma^* p}^{eff} \bf (\mu b)$}} 
\end{sideways}  
\epsfig{file=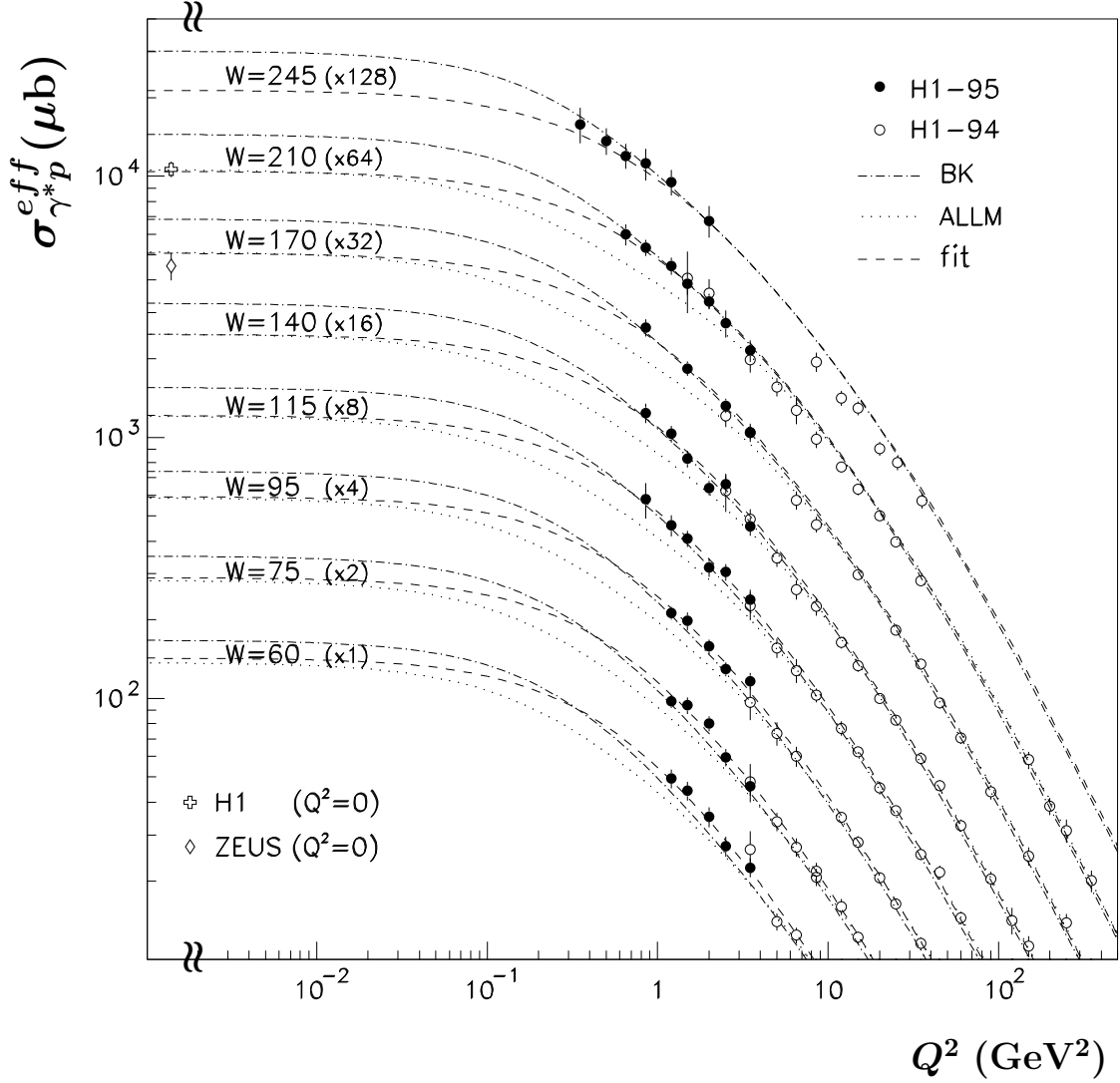,width=15.5cm,bbllx=25pt,bblly=140pt,bburx=550pt,bbury=695pt}                              
\end{center}                                                               
\begin{picture}(0,0) \put(120,15){{\Large \boldmath  $Q^2$ \bf (GeV$^2$)}} 
\put( 22.7,28.5){\LARGE \boldmath $\wr$}
\put( 24,28.5){\LARGE \boldmath $\wr$}
\put( 22.7, 154){\LARGE \boldmath $\wr$}
\put( 24, 154){\LARGE \boldmath  $\wr$}
\end{picture} 
  \caption[]{\label{fig11}                                                    
\sl  Measurement of the virtual photon-proton cross section
 $\sigma_{\gamma^*p}^{eff}$
    as a function of $Q^2$ at various values of   $W$ (in GeV). 
The cross sections for consecutive $W$ values are multiplied
with the factors indicated in the figure (numbers in brackets).
The  errors               
    represent the statistical and systematic errors added                     
    in quadrature. Full symbols are  data from this analysis;
open symbols are  H1 data from \protect \cite{F2-h1,FL-h1}.
The photoproduction points (cross: $W= 210 $ GeV, diamond:
$W= 170$ GeV) are from~\cite{H1STOT,zeustot}.
Global normalization uncertainties 
are not included in the 
errors shown.
The curves represent the  
 ALLM (dotted line)  and BK
(dashed-dotted line) 
parameterizations and the H1 fit based on BK (dashed line).}
\end{figure}
\newpage

\end{document}